\documentclass{webofc}
\usepackage[varg]{txfonts}   
\woctitle{Physics at the Magnetospheric Boundary}
\begin{document}
\title{Accretion in supergiant High Mass X-ray Binaries}
%
%

\author{Antonios Manousakis\inst{1,2}\fnsep\thanks{\email{antonism@camk.edu.pl}} \and
        Roland Walter \inst{2} \and
        John Blondin \inst{3}
}

\institute{ 
Centrum Astronomiczne im. M. Kopernika, Bartycka 18, PL-00716 Warszawa, Poland
\and
ISDC \& Observatoire de Gen\`eve, Universit\'e de Gen\`eve,  ch. d'Ecogia 16, CH-1290 Versoix, Switzerland
\and
Department of Physics, North Carolina State University, Raleigh, NC 27695-8202, USA
          }

\abstract{%
Supergiant High Mass X-ray Binary systems (sgHMXBs) consist of a massive, late type, star and a neutron star. The massive stars exhibits strong, radiatively driven, stellar winds. Wind accretion onto compact object triggers X-ray emission, which alters the stellar wind significantly. Hydrodynamic simulation has been used to study the neutron star - stellar wind interaction it two sgHMXBs: i) A heavily obscured sgHMXB (IGR $J17252-3616$) discovered by {\it INTEGRAL}. To account for observable quantities (i.e., absorbing column density) we have to assume a very slow wind terminal velocity of about 500 km/s and a rather massive neutron star. If confirmed in other obscured systems, this could provide a completely new stellar wind diagnostics. ii) A classical sgHMXB (Vela X-1) has been studied in depth to understand the origin of the off-states observed in this system. Among many models used to account for this observed behavior (clumpy wind, gating mechanism) we propose that self-organized criticality of the accretion stream is the likely reason for the observed behavior. In conclusion, the neutron star, in these two examples, acts very efficiently as a probe to study stellar winds.
}
\maketitle
\section{Introduction}
\label{intro}

In  supergiant High Mass X-ray Binaries (sgHMXBs)  neutron stars are 
orbiting at a close distance of $\alpha\sim 1.5-2$ R$_{*}$ from their 
companion stars.  The donors, in these systems,  are  OB supergiants  with 
mass loss rates of the order of $\sim10^{-6}\, M_{\odot}$ yr$^{-1}$ and  
wind terminal velocities of $\sim$ 1500 km s$^{-1}$. The neutron star accretes gas from the stellar wind and a fraction of the gravitational potential energy is converted into X-rays, ionizing and heating the surroundings. The X-ray emission can be used to investigate the structure of the stellar wind \emph{in situ} \cite{Walter07winds}.

The interactions between the neutron star and the stellar wind in Vela X-1  
 revealed that the wind of the massive star is heavily disrupted by the gravity 
 and photoionization of  the neutron star \cite{Blondin90,Blondin91}.
The  heavily obscured sgHMXBs share some of the  characteristics of the classical sgHMXBs. The main difference between classical and obscured sgHMXBs is that
 the latter ones are much more absorbed in the X-rays ($N_{H} > 10^{23}$ cm$^{-2}$)  on average, 10 times larger than in  classical systems 
and well above the galactic absorption. 

\section{Simulations \& Results}
\label{sec-2}

\subsection{The simulations}

The VH1\footnote{
http://wonka.physics.ncsu.edu/pub/VH-1/} hydrodynamical code \cite{Blondin90,Blondin91} has been employed   in order to study the interplay between the stellar wind and the compact object and compare it with the observational
features of two supergiant sgHMXB systems \cite{Manousakis+12,Manousakis+13}. Namely, to explain the origin of the obscuration of the heavily obscured eclipsing sgHMXB 
IGR $J$ 17252-3616 and to understand the origin of the X-ray off-states in the 
 classical sgHMXB Vela X-1. 
In our simulations of sgHMXBs we account for: i) the gravity of 
the primary and of the neutron star, ii) the radiative acceleration of the 
stellar wind of the primary star, and iii) 
the suppression of the stellar wind acceleration in the Str\"{o}mgren sphere of 
the neutron star.
The simulations take place  in the orbital plane, reducing the problem  to two 
dimensions. In both cases we have assumed circular orbits ($e=0$).

 The code produces  density and ionization ($\xi= L_{X}/n r_{ns}^2$, 
 where $L_{X}$ is the average X-ray luminosity, 
 $n$ is the number density at the distance $r_{ns}$
 from the neutron star \cite{1969ApJ...156..943T}) maps that are stored. 
These allow to determine the  simulated column density. 
As short time-scale variations occurs, we have 
calculated the time-averaged  orbital phase resolved column density. The 3-D instantaneous mass accretion rate ($\dot{M}_{acc}$) onto the neutron star is also recorded. From the mass accretion rate, we can infer the instantaneous X-ray luminosity of the neutron star. The relaxation time is of the order of 0.5 orbits. The first couple of days of the simulations are therefore excluded from our 
  analyses. In both simulations, 
the grid is centered on the center of mass. 
The boundary conditions are set in a way that  
matter is removed when it reached the cell including the neutron star. 
The boundary conditions at the radial outermost part of the mesh are characterized as outflow. 

 The  winds  of massive supergiant stars are  radiatively driven by absorbing 
 photons from the underlying photosphere, as described in CAK model \cite{CAKwind}. 
 However, regions in the stellar wind can differ from the predictions of the  CAK/Sobolev approximation when instabilities are 
 taken into account \citep{Owocki+88}. 
 The velocity  is  described by the  $\beta$-velocity law,
$ \upsilon=\upsilon_{\infty}(1-R_{*}/r)^{\,\beta}$
 where $\upsilon_{\infty}$ is the terminal velocity and 
 $\beta$ is the gradient of the velocity field. 
For supergiant stars,  values for wind terminal velocities 
and mass-loss rates are in the range $\upsilon_{\infty}\sim 1500-3000$ km s$^{-1}$ and
$\dot{M}_{\rm w}\sim 10^{-(6-7)}$ M$_{\odot}$ yr$^{-1}$, respectively \citep{winds_from_hot_stars}.

  A critical ionization parameter, above which the radiative force 
is negligible is defined. For $\xi>10^{2.5}$ erg cm sec$^{-1}$, most of the 
elements  responsible for the wind acceleration are  fully ionized and hence 
the radiative acceleration force vanishes. 
The main effects of the ionization is the reduction of the wind velocity in the vicinity of the neutron star and the enhancement of the mass accretion rate
 onto the compact object.

\subsection{Obscured sgHMXB: IGR {\it J}17252-3616}

The obscured sgHMXB IGR $J17252-3616$ 
 is an eclipsing binary hosting a  pulsar with P$_{s}\sim$ 414 sec, an orbital 
 period of P$_{o}\sim$ 9.74 days, and an orbital  radius of  $\alpha\approx$1.75 $
R_{*}$. 
Ground-based observations \cite{Mason_et_al09}  showed that the 
donor star is likely a B0-5I or B0-1 Ia. 
Optical and IR observations confirmed the supergiant nature and  showed 
prominent P-Cygni profile \cite{Chaty_et_al08}.
To explain the $XMM-Newton$ and $INTEGRAL$ observations, we \cite[MW11]{Manousakis+11}  suggested  that the wind terminal velocity of the system is 
relatively, low, of the order  of $\upsilon_{\infty}\sim 500$ km s$^{-1}$. 

In order to study this system, 
a computational mesh of 600 radial by 247 angular zones, 
extending  from 1 to $\sim$ 15 R$_{*}$ and in 
angle from $-\pi$ to $+\pi$, has been employed. The grid structure is 
non-uniform, with cells size decreasing towards the neutron star.
The maximal resolution, reached close to the neutron star, is $\delta R\sim 10^{10}$ cm $\sim r_{acc}/3$

\begin{figure}
\centering
\includegraphics[width=0.495\textwidth,clip]{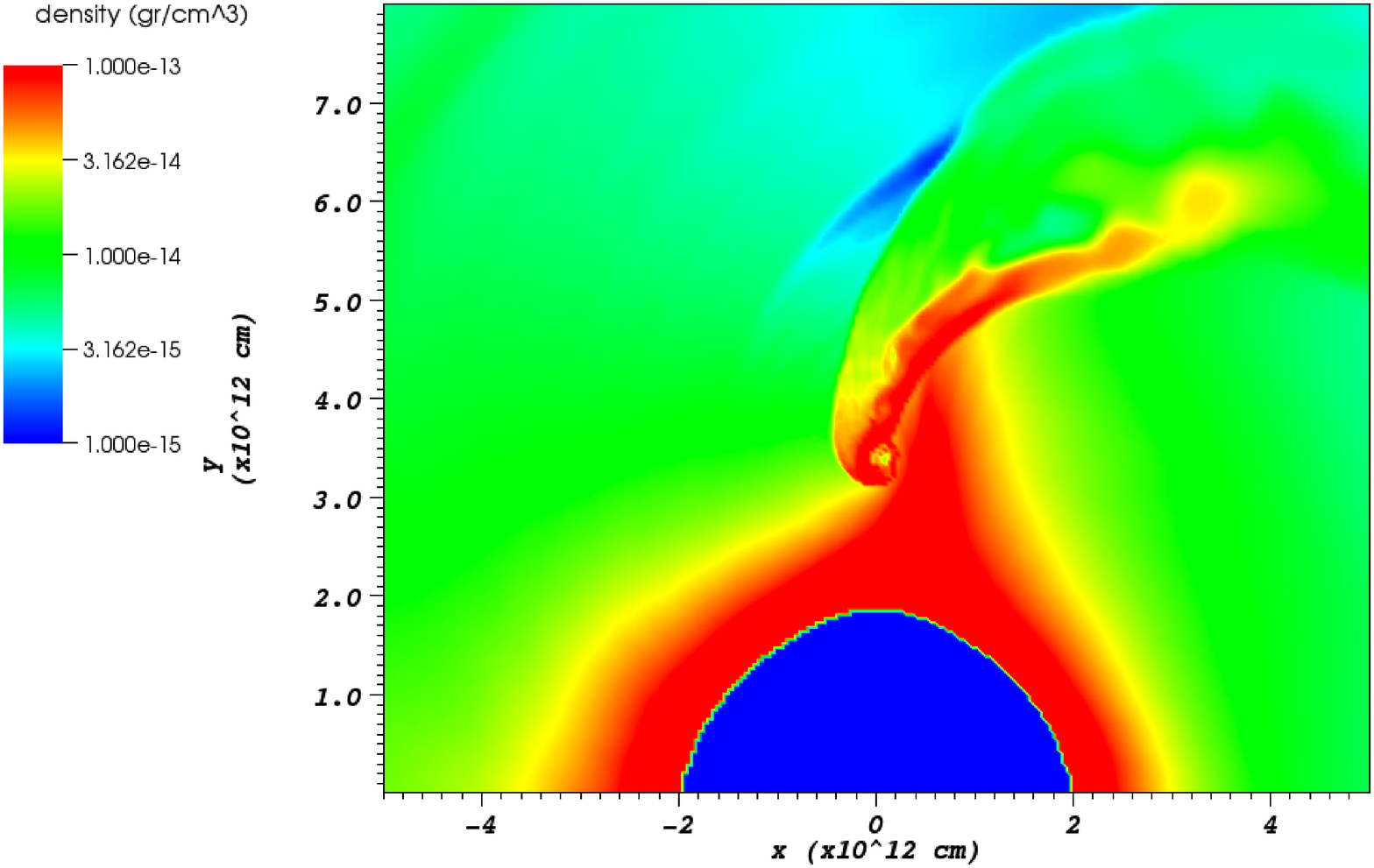}
\hspace{-0.15cm}
\includegraphics[width=0.495\textwidth,clip]{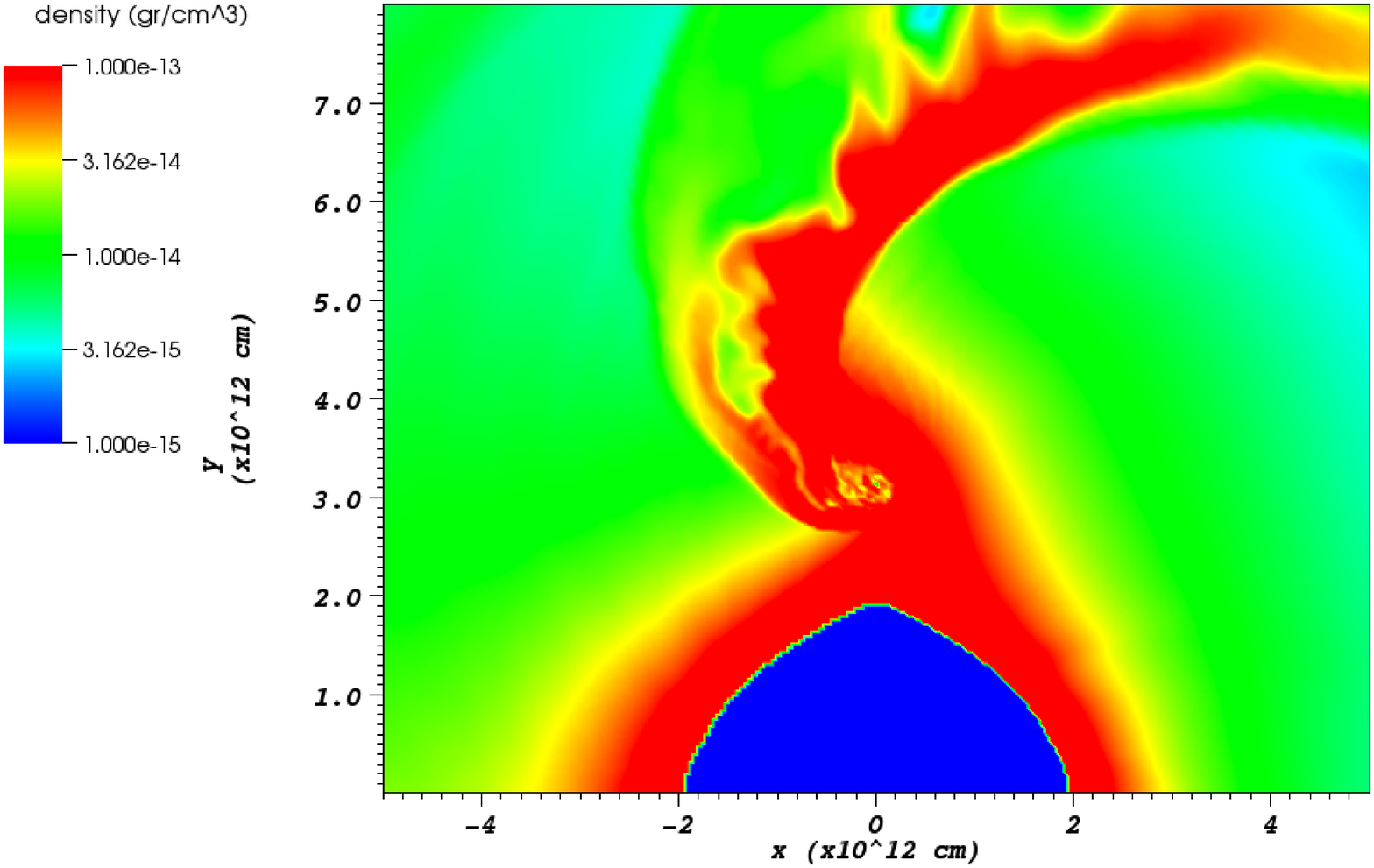}
\caption{Density distribution (in gr cm$^{-3}$; color bar) on the orbital plane 
after $\sim$ 3 orbits. The wind terminal velocity is $\upsilon_{\infty}\approx 
500$ km s$^{-1}$ and the mass-loss rate is $\dot{M}_{w}\approx 10^{-6}\, M_{\odot}$ y
r$^{-1}$. The mass of the neutron star scales
  from 1.5 (left) to 2.0 $M_{\odot}$ (right). 
  The color version of this figure is available on-line.
}
\label{fig-1}       
\end{figure}

The appropriate modeling of stellar wind  resulted in a smooth stellar wind with a 
 terminal velocity of $\upsilon_{\infty}\approx 
500$ km s$^{-1}$ and a mass loss rate of $\dot{M}_{w}\approx 10^{-6}\, M_{\odot}$ y
r$^{-1}$. This wind terminal velocity is about 2-3 times less 
than expected for an OB supergiant of the same spectral type and 
effective temperature.
The effect of the gravity of the neutron star deflecting the accretion flow as can be seen in figure \ref{fig-1}  between the left (for a neutron star mass of 1.5 M$_{\odot}$) and the 
right (for a neutron star mass of 2.0 M$_{\odot}$) panels.  
For a heavier neutron star, more gas will accumulate and the  
absorption will get stronger.

\begin{figure}
\centering
\includegraphics[width=0.4\textwidth,clip]{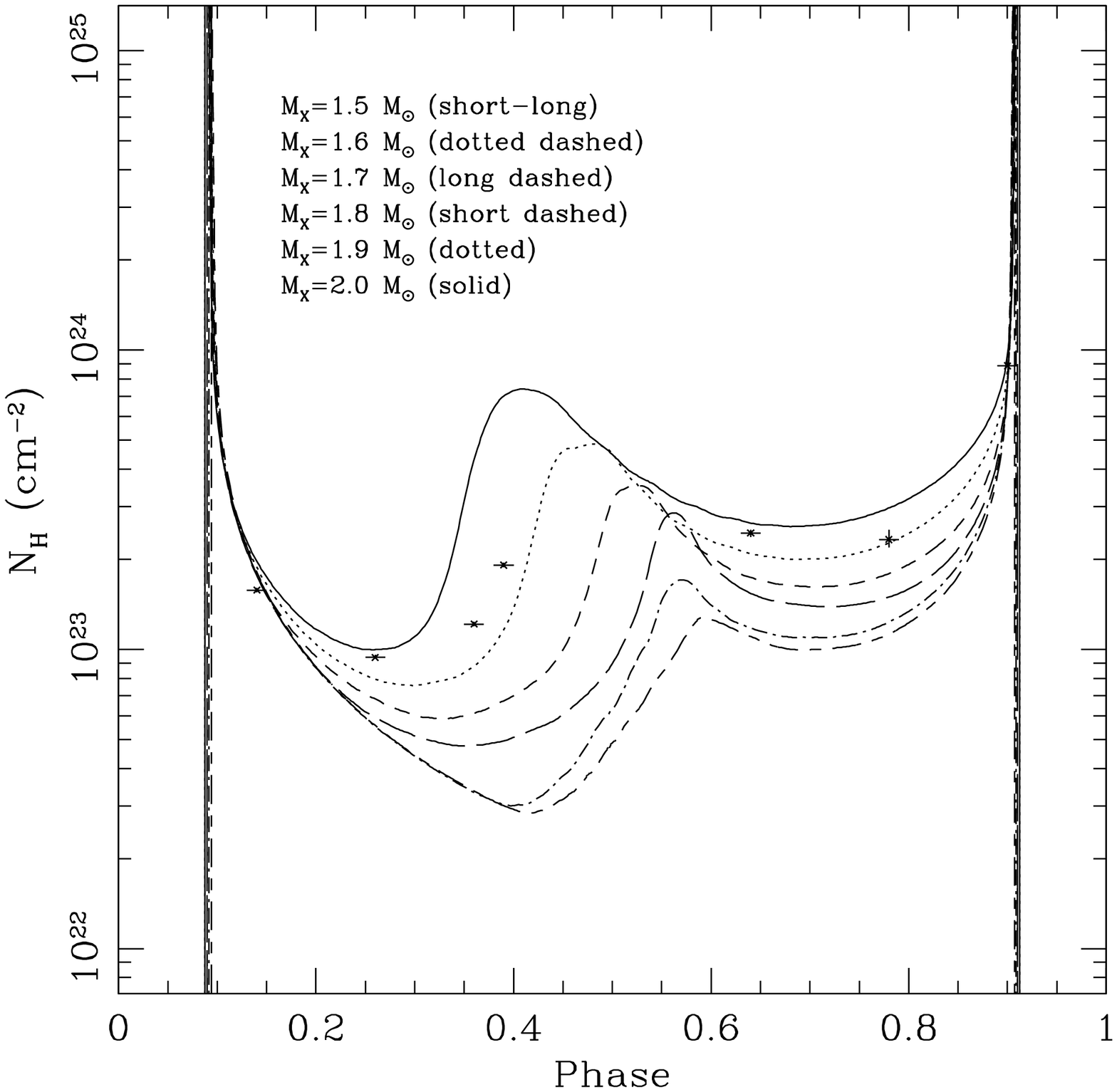}
\includegraphics[width=0.55\textwidth,clip]{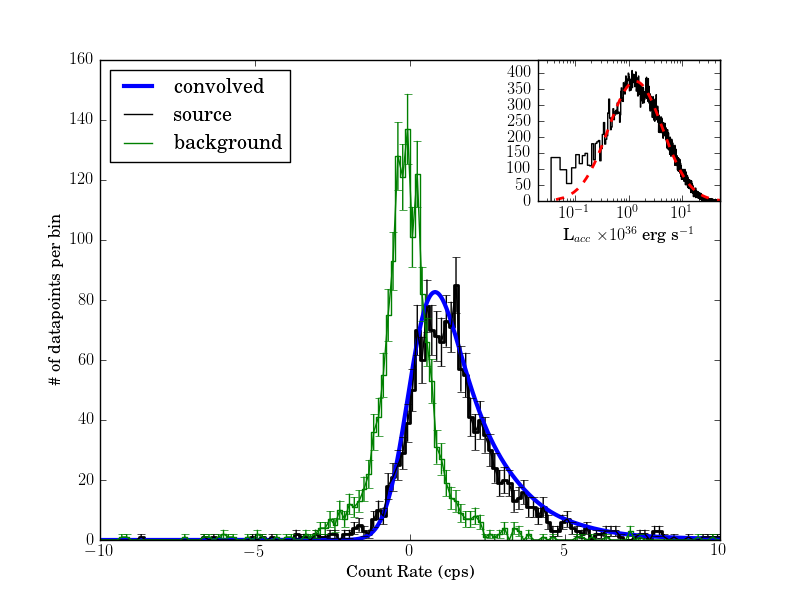}
\caption{{\it Left:} Time-averaged simulated absorbing column density 
($N_{H}$) for various neutron star masses. The two limiting cases (short-long dashed and solid) are the results from figure \ref{fig-1}. 
The data points are  from MW11. 
{\it Right:} The histogram of observed light-curve  (black), 
the background fluctuation (green), and the convolution of the simulated 
 data with the latter (blue). The inset shows the corresponding distribution of the simulated. light-curve (black) with  a log-normal fit (red dashed curve) 
  The color version of this figure is available on-line.
}
\label{fig-2}       
\end{figure}

The  heavier  the neutron star, the more bended the accretion wake. This effect
is  revealed in the orbital phase dependency of the absorbing column density (see
fig. \ref{fig-2}; left panel), 
moving the position of the minimum to earlier phases.   
The histogram of the source observed 
light-curve (black) as well as the background (green), obtained with INTEGRAL through HEAVENS\footnote{http://www.isdc.unige.ch/heavens/} interface are shown in figure \ref{fig-2} (right panel).  The inset panel 
shows the histogram derived from the simulated light-curve, and is well fit 
by a log-normal distribution. The convolution of the simulated data and the background is then compared with the source (blue curve). 
We can therefore estimate, in the frame of our model, that the mass of the neutron star  is in  the range $M_{NS}=1.75-2.15$.

\vspace{0.25 cm}
{\it The outcome of this work is described in Manousakis, Walter \& Blondin \cite{Manousakis+12} 
}

\subsection{The classical sgHMXB Vela X-1}

Vela X-1 (=4U 0900$-$40) is a classical eclipsing super-giant High Mass X-ray 
Binary (sgHMXB). The system consists of an evolved B 0.5 Ib supergiant (HD77581) and of a massive neutron star of M$_{NS}=1.86$ M$_{\odot}$ \cite{Quaintrell_et_al03}. 
The neutron star orbits its massive companion with a period of about 8.9 days,
 in a  circular orbit with a 
 radius of $\alpha=$1.76 $R_{*}$ \cite{1997ApJS..113..367B}. 
 The  stellar wind is characterized by a mass-loss rate of $\sim 4\times10^{-6}$ M$_{\odot}$ yr$^{-1}$ \cite{1986PASJ...38..547N} 
and a wind terminal velocity of $\upsilon_{\infty}\approx 1700$ km s$^{-1}$ 
\cite{1980ApJ...238..969D}. The X-ray luminosity is typically $\sim 4 \times 10^{36
}$ erg s$^{-1}$, although high variability can be observed.
Recent studies on the hard X-ray variability of Vela X-1 have revealed a 
rich phenomenology. Flaring activity and short off-states have been observed 
\cite{Kreykenbohm+08}. Both flaring activity and off-states were interpreted 
as the effect of a strongly structured wind, characterizing the X-ray variability of Vela X-1 with a log-normal distribution, interpreted in the context of a clumpy stellar wind \cite{clumpingVela}. The quasi-spherical subsonic 
accretion model \cite{2012MNRAS.420..216S,2013MNRAS.428..670S} is an alternative, 
predicting that the repeatedly observed off-states in Vela X-1 are the
result of a transition from the Compton cooling 
(higher luminosity) to radiative cooling (lower luminosity)

A computational mesh of 900 radial by 347 angular zones, extending  
from 1 to $\sim$ 25 R$_{*}$ and in 
angle from $-\pi$ to $+\pi$, has been employed. 
The grid is built in a non-uniform way in order to allow
 for higher resolution of $\sim 10^{-9}$ cm  at the neutron star.

\begin{figure} 
\centering
\vspace{-2.5cm}
\includegraphics[width=0.75\textwidth,clip]{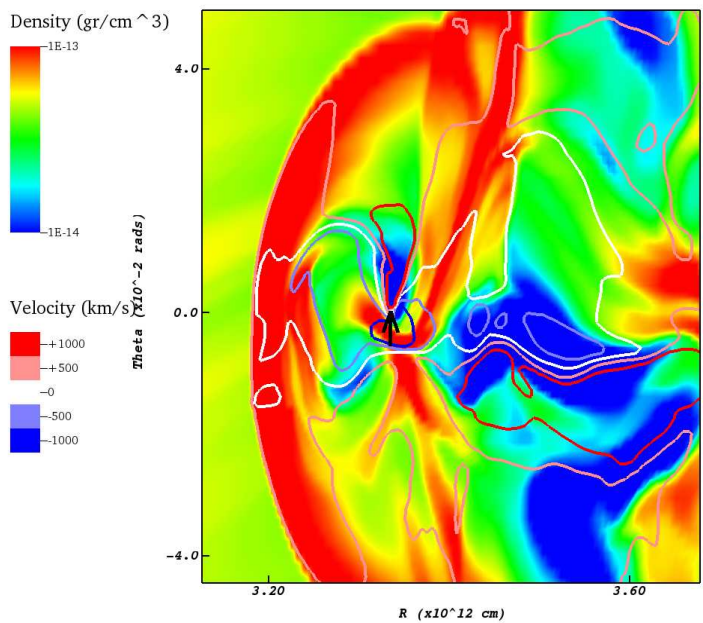}
\vspace{-3cm}
\caption{
Density distribution (in gr cm$^{-3}$)  during the off-state. The over-plotted
contours shows the radial velocities. 
The position of the neutron star is indicated by the black arrow.
  The color version of this figure is available on-line.
}
\label{fig-3}       
\end{figure}

\begin{figure} 
\centering
\includegraphics[width=0.47\textwidth,clip]{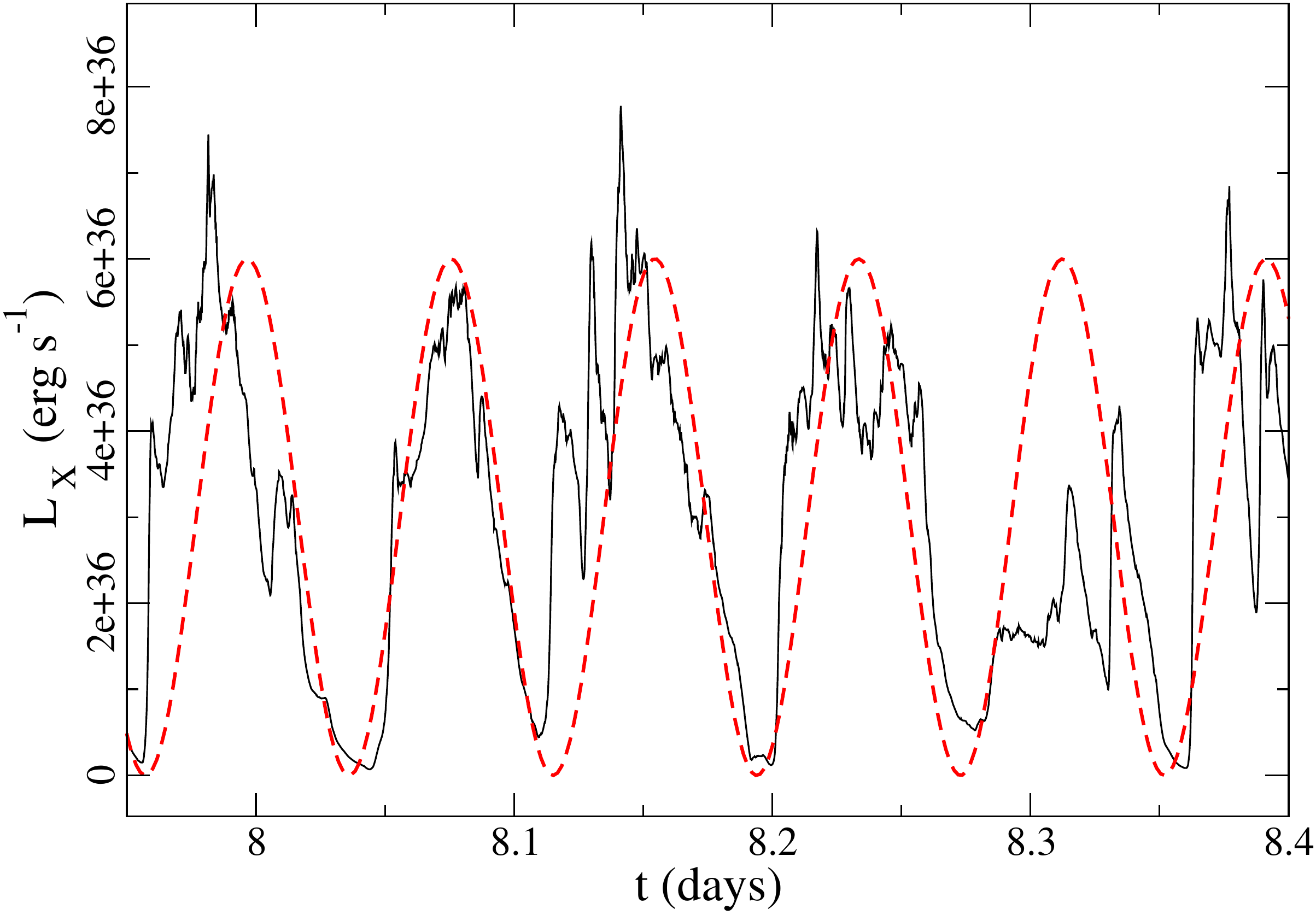}
\includegraphics[width=0.47\textwidth,clip]{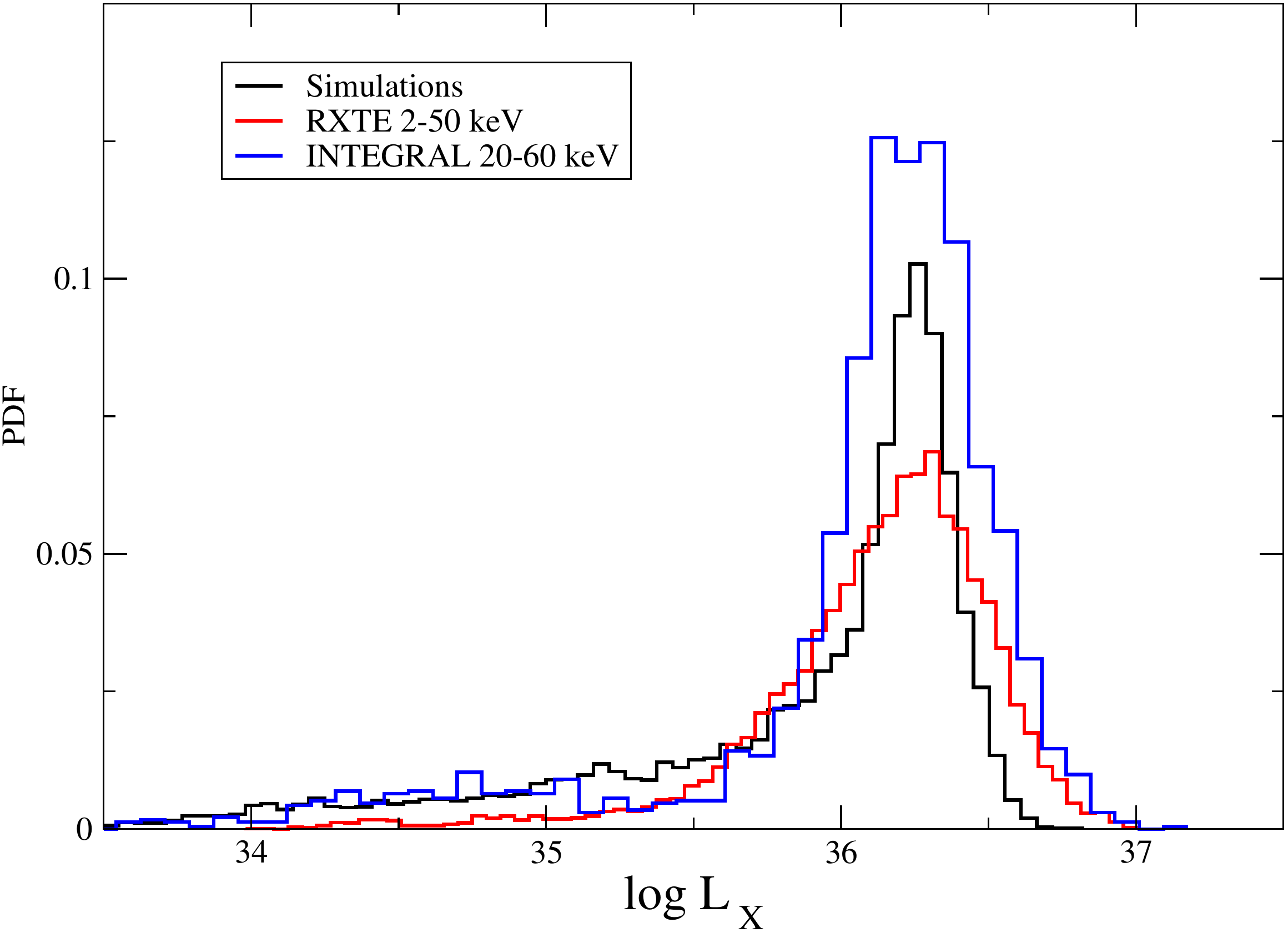}
\caption{{\it Left:} 
Simulated light-curve of Vela X-1 together with a 
sine wave of $6820$ sec period (red dashed line). 
This quasi-periodic modulations was observed in Vela X-1 by \cite{Kreykenbohm+08}.
The color version of this figure is available on-line.
{\it Right:} 
The X-ray luminosity  distribution of Vela X-1 from RXTE (2-50 keV; red) 
and INTEGRAL (20-60 keV; blue) observations together with that derived from the simulations (black). All the distributions are normalized to unity.
}
\label{fig-4}       
\end{figure}

Our simulations predicts the formation of low density
bubbles behind the bow shock, around the neutron star, resulting in 
the X-ray off-states. These bubbles are $\sim$ 10 times
larger than the accretion radius. When a bow shock 
appears, it moves away from the neutron star up to a 
distance of $\sim$ 10$^{11}$ cm (see fig. \ref{fig-3}). A low density bubble 
forms behind and starts expanding. It then gradually fall back and a
stream of gas eventually reaches the neutron star and produces a new rise 
of the X-ray flux. The accretion stream can either move left-handed 
or right-handed. This ‘breathing’ behavior is quasi-periodic.
This behavior is illustrated in figure \ref{fig-4} (left panel) with the 
simulated X-ray lightcurve over-plotted with a sinusoidal curve with a 
period of 6820 sec. The observed light-curves also show a 
quasi-periodic signal at $\sim$ 6800 sec \cite{Kreykenbohm+08}. This
modulation is related to the characteristic free-fall time of
the low density bubble (radius of $\sim 10^{11}$ cm) much longer
than the accretion or magnetospheric radii.

We have constructed histograms of the observed and simulated
light-curves (see fig. \ref{fig-4}; right panel). The histograms are 
normalized to an integral equals to unity. All three distribution can 
be fit with the log-normal distribution characterized by a 
standard deviation ($\sigma$) of 0.23 for INTEGRAL (blue) and  0.30 for RXTE (red), while 
the distribution is narrower for simulated lightcurve (black) having  a standard deviations of 0.2. A minor excess can be seen at the lower end of the 
distribution at about $L_{X}\sim 10^{35}$ erg s$^{-1}$.

The hydrodynamic simulations of Vela X-1 are 
sufficient to explain the observed behavior without the need 
for clumpy stellar wind or high magnetic fields and gating 
mechanisms. Self-organized criticality \cite{1988PhRvA..38..364B} of the 
accretion stream is enough to describe the observed variability.

\vspace{0.25 cm}
{\it The outcome of this work is described in Manousakis \& Walter \cite{Manousakis+13}
}

\section{Conclusion}

We have compared observed properties with the results of hydrodynamical 
simulations  in two sgHMXB systems. Our main conclusions are as follows, 

\begin{itemize}
\item The obscured sgHMXBs can bee understood in terms of low wind 
terminal velocities. 

\item The comparison between the hydrodynamical simulations with  
observations, in obscured sgHMXB, 
allow to constrain neutron star mass (and orbital radius), 
independently of dynamical estimates. 

\item In both simulations, we are able to produce log-normal distributions of the 
accretion rates. In addition, for Vela X-1 we are able to produce the X-ray off-states.

\item Self-organized criticality of the accretion stream can explain the observed 
variability. 

\end{itemize}

\bibliography{references}

\end{document}